\documentclass[aps,pra,twocolumn]{revtex4}
\usepackage{graphicx}

\begin{document}

\title{Phase diagram of a polarized Fermi gas across a Feshbach resonance in a potential trap}
\author{W. Yi and L.-M. Duan}
\address{FOCUS center and MCTP, Department of Physics, University of Michigan, Ann
Arbor, MI 48109}

\begin{abstract}
We map out the detailed phase diagram of a trapped ultracold Fermi
gas with population imbalance across a wide Feshbach resonance. We
show that under the local density approximation, the properties of
the atoms in any (anisotropic) harmonic traps are universally
characterized by three dimensionless parameters: the normalized
temperature, the dimensionless interaction strength, and the
population imbalance. We then discuss the possible quantum phases in
the trap, and quantitatively characterize their phase boundaries in
various typical parameter regions.
\end{abstract}

\maketitle

\section{Introduction}

Recent experiments with ultracold Fermi gases near a Feshbach
resonance, through which the inter-atomic interaction can be tuned
by varying the external magnetic field, have attracted
considerable attention \cite{1}. The latest advance in this
direction features two very recent experiments \cite {2,3}, which
study resonantly interacting ultracold ${}^{6}$Li atoms in a trap
with a population imbalance of different spin components. The
pairing superfluidity in the Fermi gases with population imbalance
between the two spin components have been studied before in
different physical contexts, mostly in the weakly interacting
regime \cite{4,5,6,7,8,8b,8a}. Some exotic phases have been
proposed to exist due to the competition between the Cooper
pairing and the population imbalance \cite{5,7,8,8a}. The recent
experiments near a Feshbach resonance have raised strong
theoretical interest in studying the phase configuration of such a
system under a potential trap in the strongly interacting region
\cite{9a,9b,9c,9d,9e,9f,9g,9h}.

In this work, we map out the detailed phase diagram for fermionic
atoms in a trap with population imbalance, both at zero and at
finite temperature. We use the same theoretical method as proposed
in Ref. \cite{9c}, which corresponds to a generalization of the
self-consistent $G_{0}G$ diagram scheme \cite{10} from the equal
population case to the case with population imbalance. At zero
temperature, this method reduces to the mean-field approximation
for the crossover theory \cite{10a,9c,9e,9f}; while at finite
temperature, it includes a pseudogap in addition to the superfluid
order parameter. To avoid subtle unstable solutions for the ground
state of this system, we directly minimize the thermodynamic
potential under the local density approximation instead of using
the gap equation.

One of the difficulties to map out the detailed phase diagram lies
in the fact that the properties of the system seem to depend on so
many different parameters. For instance, we expect in general,
several different phases to exist from the trap center to the
edge, with their characters and boundaries determined by the
temperature of the system, the population imbalance, the magnetic
field detuning, the atom specie, the trap frequencies along the
three spatial directions, and the total atom number. It is
difficult to calculate the phase distribution for all these
different parameters. Fortunately, similar to the homogeneous
system with equal spin populations, there exists a nice
universality for this more involved system. Although the
properties of the system depend on all the parameters mentioned
above, the dependence is through some dimensionless combinations
of the physical parameters. As a result, the phase diagram is
completely fixed by three dimensionless parameters after
rescaling: the normalized temperature, the dimensionless
interaction strength, and the population imbalance. In particular,
at zero temperature and at the resonance point, the phase diagram
only depends on a single parameter: the population imbalance. In
this universal frame, the variations in the trap (anisotropic in
general) or in the atom number do not induce any further
complexity for the description of the system.

To fix the phase diagram, we calculate under various interaction
strengths and temperatures, the distribution of the system's phases
from the trap center to the edge as a function of the population
imbalance. The main results are shown in Fig. 1 and 3. In the
following, we first give the universal equations of state in Sec.
II, written in terms of the three dimensionless parameters. In Sec.
III and IV, we present our main calculation results with detailed
discussions.

\section{The Formalism for a Trapped Fermi Gas with Population Imbalance and
the Universality}

As the population of the closed channel molecules is exceedingly small near
a wide Feshbach resonance \cite{10,11}, it is sufficient to use the
following single-channel Hamiltonian to describe the Fermi gas in the near
resonance region:
\begin{eqnarray}
H &=&\sum_{\mathbf{k},\sigma }(\epsilon _{\mathbf{k}}-\mu _{\sigma })a_{%
\mathbf{k},\sigma }^{\dag }a_{\mathbf{k},\sigma }   \\
&+&\left( U/\mathcal{V}\right) \sum_{\mathbf{q},\mathbf{k},\mathbf{k^{\prime
}}}a_{\mathbf{q}/2+\mathbf{k},\uparrow }^{\dag }a_{\mathbf{q}/2-\mathbf{k}%
,\downarrow }^{\dag }a_{\mathbf{q}/2-\mathbf{k^{\prime }},\downarrow }a_{%
\mathbf{q}/2+\mathbf{k^{\prime }},\uparrow }\nonumber
\end{eqnarray}
where $\epsilon _{\mathbf{k}}=k^{2}/(2m)$ ($m$ is the atom mass and $\hbar =1
$), $\mu _{\sigma }$ is the chemical potential for the spin-$\sigma $
component ($\sigma =\uparrow ,\downarrow $ labels the two spin states), $%
\mathcal{V}$ is the quantization volume, $a_{\mathbf{k},\sigma
}^{\dag }$ is the creation operator for the fermionic atoms. The
bare atom-atom interaction rate $U$ is connected with the physical
one $U_{p}=4\pi a_{s}/m$ ($a_{s}$ is the atomic scattering length)
through the standard renormalization relation $1/U=1/U_{p}-\left(
1/\mathcal{V}\right) \sum_{\mathbf{k}}1/\left( 2\epsilon
_{\mathbf{k}}\right) $ \cite{10}. We take the local density
approximation so that $\mu _{\uparrow }=\mu _{\mathbf{r}}+h,$ $\mu
_{\downarrow }=\mu _{\mathbf{r}}-h$, $\mu _{\mathbf{r}}=\mu -V\left( \mathbf{%
r}\right) $, where $V\left( \mathbf{r}\right) $ is the external trap
potential (slowly varying in $\mathbf{r}$). The chemical potential $\mu $ at
the trap center and the chemical potential imbalance $h$ are determined from
the total atom number $N=N_{\uparrow }+N_{\downarrow }$ and the population
imbalance $\beta =\left| N_{\uparrow }-N_{\downarrow }\right| /N$ through
the number equations below.

As has been shown in Ref. \cite{9c}, under the local density
approximation, the thermodynamic potential $\Omega =-T\ln
[$tr$\left( e^{-H/T}\right) ]$ corresponding to Hamiltonian (1) has
the following expression:
\begin{eqnarray}
\Omega /\mathcal{V}&=&-|\Delta| ^{2}/U_p-\left(
T/\mathcal{V}\right) \sum_{ \mathbf{k}}\{\ln [1+\exp (-\left|
E_{\mathbf{k}\downarrow }\right| /T)] \nonumber\\&&+\ln[1+\exp
(-\left| E_{\mathbf{k}\uparrow }\right|
/T)]-|\Delta|^2/(2\epsilon_{\mathbf{k}}T)\\&& -[\epsilon _{
\mathbf{k}}-\mu _{\downarrow }-\theta (E_{\mathbf{k}\downarrow
})E_{\mathbf{k }\downarrow }+\theta (-E_{\mathbf{k}\uparrow
})E_{\mathbf{k}\uparrow }]/T\nonumber \},
\end{eqnarray}
where the gap $\Delta $ at zero temperature is given by the order parameter $%
\Delta _{s}=U\sum_{\mathbf{k}}\langle a_{\mathbf{-k},\downarrow }a_{\mathbf{k%
},\uparrow }\rangle $; and at finite temperature should be
understood as the total gap, with $|\Delta| =\sqrt{\left| \Delta
_{s}\right| ^{2}+\left| \Delta _{pg}\right| ^{2}}$, where $\Delta
_{pg}$ is the pseudogap coming from the contribution of
non-condensed Cooper pairs \cite{10}. The $%
\theta $-function is defined as $\theta \left( x\right) =1$ for $x>0$ and $%
\theta \left( x\right) =0$ otherwise. Without loss of generality, we take $%
h>0$ so that $N_{\uparrow }>N_{\downarrow }$ always. Note that different
from the equal-population case, the quasi-particle excitation energies $E_{\mathbf{k\sigma }%
}$ are different for the $\mathbf{\sigma =}\uparrow ,\downarrow $
branches: $E_{\mathbf{k}\uparrow ,\downarrow }=\sqrt{(\epsilon _{%
\mathbf{k}}-\mu _{\mathbf{r}})^{2}+\left| \Delta \right| ^{2}}\mp
h$. In the case of $h>0$, $E_{\mathbf{k}\downarrow }$ is always
positive; while
there exists certain parameter regions where the sign of $E_{\mathbf{k}%
\uparrow }$ becomes momentum dependent and is negative in the range $%
k_{-}<\left| \mathbf{k}\right| <k_{+}$, where $k_{\pm }^{2}=\max
\left[ 0,2m(\mu _{r}\pm \sqrt{h^{2}-|\Delta| ^{2}})\right] $. In
this momentum range, the atoms stay unpaired, which corresponds to
the so-called breached pair state \cite{7,note3}. In deriving the
thermodynamic potential (2), we have neglected the
non-zero-momentum
pairing (the FFLO state \cite{5}, with the pair momentum $\mathbf{q\neq }0$%
). This is motivated by the fact that the FFLO state is stable only within a
narrow parameter window in the BCS region \cite{8a,9b,note1} and is absent
in the recent $^{6}$Li experiments \cite{2,3}.

From the thermodynamic potential, one can get the gap equation
from the condition $\partial \Omega /\partial \Delta =0$. However,
as discussed in Ref. \cite{9c}, in the presence of a population
imbalance, the thermodynamic potential has a double well
structure, and the gap equation may give unstable solutions.
Therefore, instead of solving the gap equation, we directly
minimize the thermodynamic potential $\Omega $ to find its global
minimum with respect to the gap parameter $\Delta $. To fix the
chemical potentials $\mu _{\sigma }$ in Eq. (2), we need to use
the number equations, derived from the relations $\partial \Omega
/\partial \mu _{\sigma }=-n_{\mathbf{r}\sigma }\mathcal{V}$, where
$n_{\mathbf{r}\sigma }$ is the number density of the spin-$\sigma
$ component at the position $\mathbf{r}$. The number equations
have the form
\begin{equation}
n_{\mathbf{r}\sigma }=\frac{1}{\mathcal{V}}\sum_{\mathbf{k}}[u_{\mathbf{k}%
}^{2}f(E_{\mathbf{k},\sigma })+v_{\mathbf{k}}^{2}f(-E_{\mathbf{k},-\sigma
})],
\end{equation}
where the parameters $u_{\mathbf{k}}^{2}=(E_{\mathbf{k}}+(\epsilon _{\mathbf{%
k}}-\mu _{\mathbf{r}}))/2E_{\mathbf{k}}$, $v_{\mathbf{k}}^{2}=(E_{\mathbf{k}%
}-(\epsilon _{\mathbf{k}}-\mu _{\mathbf{r}}))/2E_{\mathbf{k}}$, $E_{\mathbf{k%
}}=\sqrt{(\epsilon _{\mathbf{k}}-\mu _{\mathbf{r}})^{2}+\left|
\Delta \right| ^{2}}$, the Fermi distribution $f(E)\equiv 1/\left(
1+e^{E/T}\right) $, and for convenience, we take $-\uparrow
=\downarrow $ and vice
versa. The atom densities $n_{\mathbf{r}\uparrow }$ and $n_{\mathbf{r}%
\downarrow }$ are connected with the total atom number and the population
imbalance through $N=\int d^{3}\mathbf{r}\left( n_{\mathbf{r}\uparrow }+n_{%
\mathbf{r}\downarrow }\right) $, and $\beta =\int d^{3}\mathbf{r}\delta n_{%
\mathbf{r}}/N$ ($\delta n_{\mathbf{r}}\equiv n_{\mathbf{r}\uparrow }-n_{%
\mathbf{r}\downarrow }$).

Given an external potential $V\left( \mathbf{r}\right) $, we can
determine the properties of the system from Eqs. (2) and (3).
However, the solution in general depends on too many external
parameters.
If the trapping potential is harmonic with the form $V(\mathbf{r})=\sum_{i}%
\frac{1}{2}m\omega _{i}^{2}r_{i}^{2}$ ($i=x,y,z$, anisotropic in
general), the properties of the system obviously will depend on the
temperature $T$, the effective scattering length $a_{s}$, the atom
mass $m$, the total atom
number $N$, the population imbalance $\beta $, and the trap frequencies $%
\omega _{i}$ along the three spatial dimensions. This much involved
dependence can be significantly simplified if we transform the set
of equations above into the dimensionless form. For that purpose, we
choose the unit of energy to be the Fermi energy ($E_{F}$) at the
center of the trap for $N$ non-interacting fermions with equal
population for the two spin components. Under the local density
approximation, one can easily figure out $ E_{F}=(3N\omega
_{x}\omega _{y}\omega _{z})^{\frac{1}{3}}$ from its definition. The
harmonic trap in the unit of $E_{F}$ can be expressed in the
standard dimensionless form $V(\mathbf{r)}/E_{F}=\sum_{i}\widetilde{r}%
_{i}^{2}$, where $\widetilde{r}_{i}\equiv r_{i}/R_{i}$, with the
Thomas-Fermi radius $R_{i}\equiv \sqrt{2E_{F}/m\omega _{i}^{2}}$
along the $i$th direction. With these, the number equations in (3)
are cast into the following dimensionless form:
\begin{equation}
1\pm\beta=\frac{6}{\pi^3}\int d^3\widetilde{r}d^3k[u_{\mathbf{k}
}^{2}f(E_{\mathbf{k},\sigma
})+v_{\mathbf{k}}^{2}f(-E_{\mathbf{k},-\sigma })],
\end{equation}
where the energies and the momenta are in the units of $E_{F}$ and
$k_{F}$ ($k_{F}\equiv \sqrt{2mE_{F}/\hbar ^{2}}$), respectively.
The dimensionless thermodynamic potential has the same form as Eq.
(2), except that the interaction strength $U_p$ is replaced by the
dimensionless one $\frac{8}{3\pi}k_{F}a_{s}$, and that all the
energies are normalized by the chosen unit $E_{F}$
(correspondingly, $\mathbf{k}$ by $k_{F}$ and $T$ by $T_{F}\equiv
E_{F}/k_{B}$).

From these dimensionless equations, it becomes obvious that the
properties of the system depend only on the three dimensionless
parameters $ T/T_{F}$, $k_{F}a_{s}$ and $\beta $. The system will
have the same properties as long as these three parameters are the
same, whether it is for different atom species, or with different
total atom numbers, or in traps with different trapping
frequencies ($\omega _{i}$). This shows that the properties of
this inhomogeneous system still have nice universality, similar to
the case of a homogeneous Fermi gas without the population
imbalance \cite{10a,universal}, where the system can be
characterized by two dimensionless parameters $k_{F}a_{s}$ and
$T/T_{F}$. Comparing with the case of a homogeneous gas, we see
that in the current case, the presence of (anisotropic) traps with
various atom numbers do not add complexity to the description of
the system. This nice feature comes from the local density
approximation and the assumption of harmonic traps \cite{note2},
and is independent of the particular approximation schemes in
deriving the equations of state.

\section{Phase Boundaries at Zero Temperature}

Following the formalism outlined in the previous section, we first
map out the phase boundaries for trapped fermions at zero
temperature. The different phases in the trap can be identified from
the gap $\Delta $ and the chemical potentials $\mu _{\sigma }$ of
the two spin components. From the solutions of $\Delta $ and $\mu
_{\sigma }$, one can immediately get the density profiles of the
trapped atoms. There are in general four possible phases in this
system \cite{9c,9e}: (i) a superfluid (SF) state with $\Delta >0$
and no Fermi surface in the momentum space; (ii) a breached pair
(BP1) state with $\Delta
>0$, one Fermi surface in the momentum space, and gapless fermionic
excitations \cite{note3}; (iii) a normal polarized (NP) state with
$\Delta =0 $ and one Fermi surface ($\mu _{\uparrow }>0,$ $\mu
_{\downarrow }<0$); and (iv) a normal mixed (NM) state with $\Delta
=0$ and two Fermi surfaces ($\mu _{\uparrow }>\mu _{\downarrow
}>0$). At zero temperature, the density profiles of the two spin
components are identical in the SF phase, but are different in the
BP1 phase. This can serve as the measure to distinguish these two
phases at $T=0$. At finite $T$, both of the phases have finite
polarization (excess fermions) with different density profiles for
the two spin components \cite{9c}. In this case, they can only be
distinguished by the existence of the Fermi surface in the BP1 phase
(see the note \cite{note3}).

\begin{figure*}[p]
\includegraphics{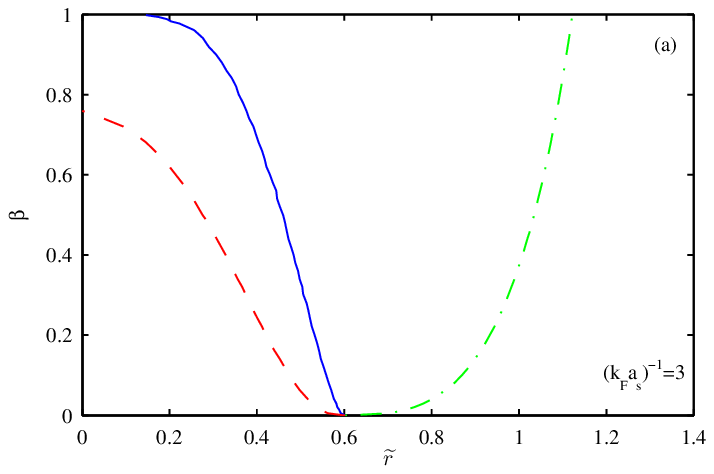} \includegraphics{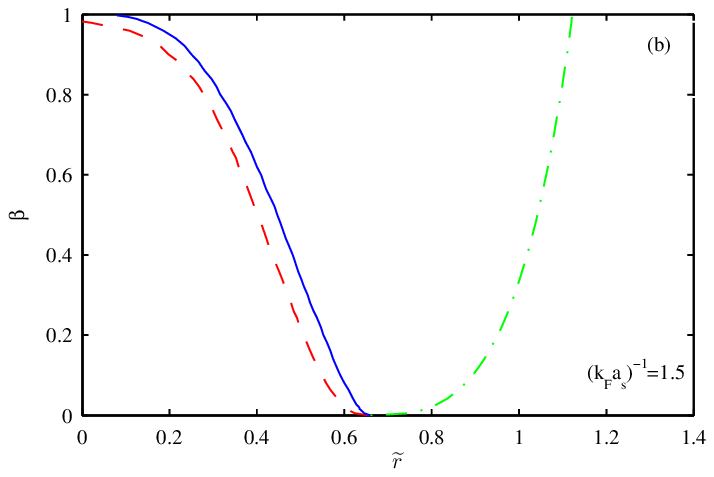} %
\includegraphics{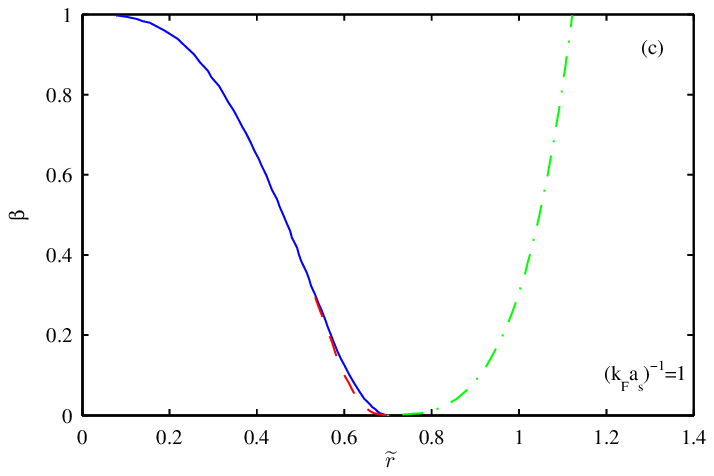} \includegraphics{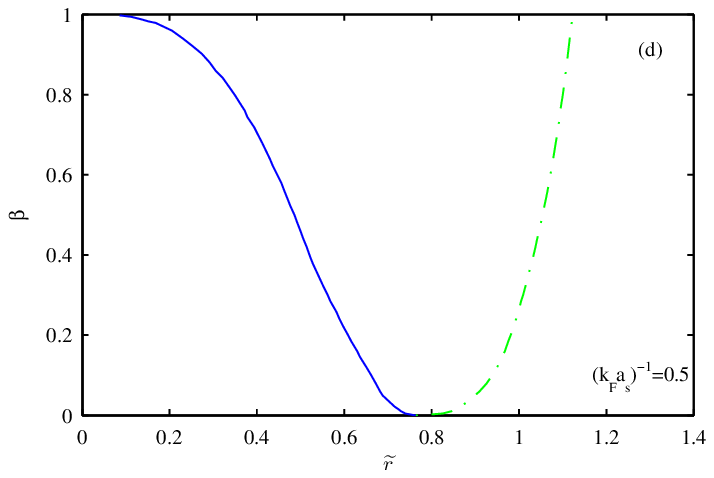} %
\includegraphics{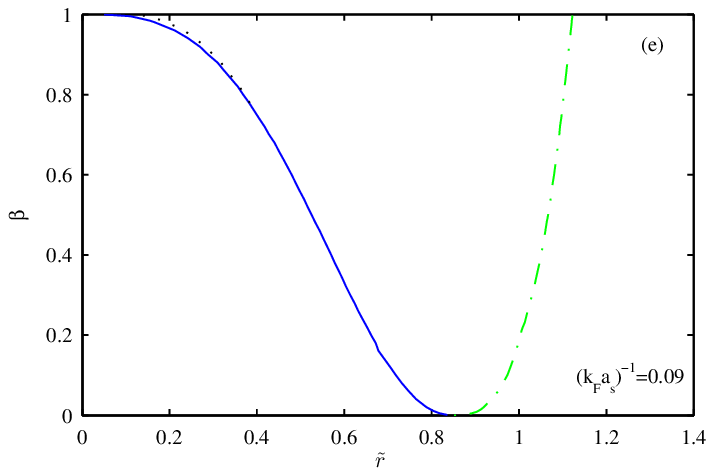} \includegraphics{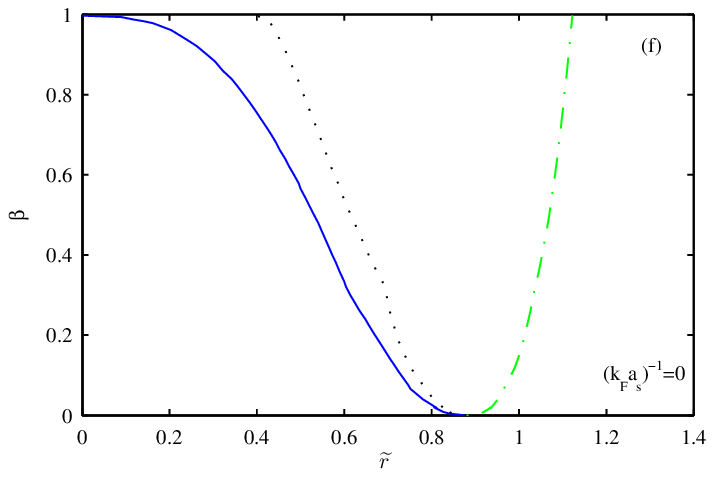} %
\includegraphics{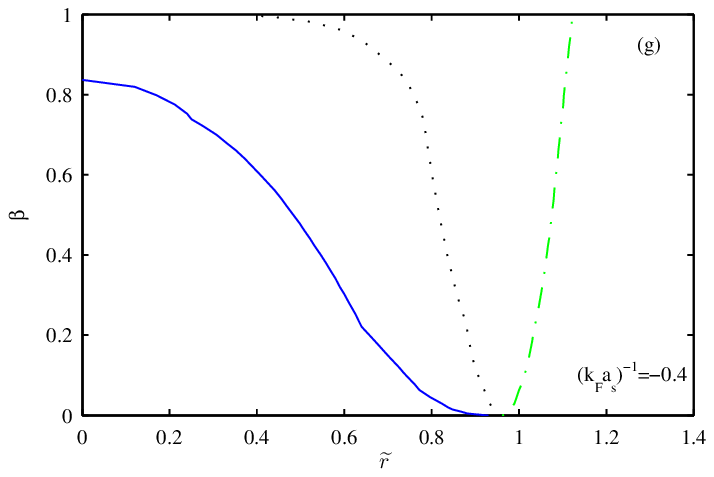} \includegraphics{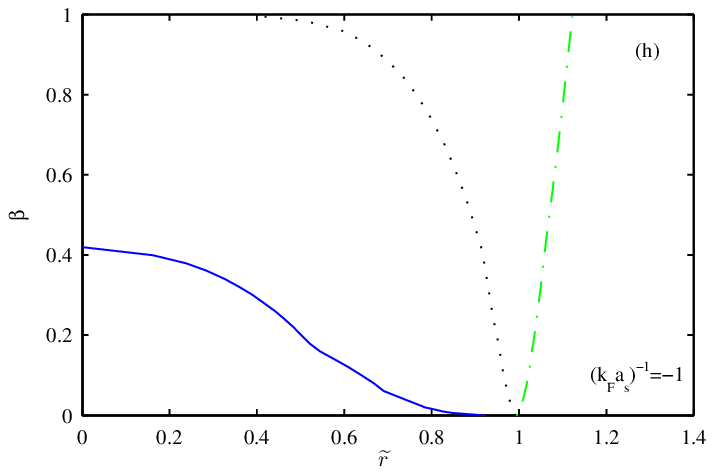}
\caption[Fig.1 ]{(Color Online) The zero temperature phase boundaries of a
polarized Fermi gas in a trap with (a)$(k_{F}a_{s})^{-1}=3$, (b) $%
(k_{F}a_{s})^{-1}=1.5$, (c) $(k_{F}a_{s})^{-1}=1$, (d) $(k_{F}a_{s})^{-1}=0.5
$, (e) $(k_{F}a_{s})^{-1}=0.09$, (f) $(k_{F}a_{s})^{-1}=0$, (g) $%
(k_{F}a_{s})^{-1}=-0.4$, (h) $(k_Fa_{s})^{-1}=-1$. The solid lines
mark the phase boundaries between the superfluid region (SF/BP1) and
the normal region (NM/NP); the dashed lines are the phase boundaries
between the SF phase and the BP1 phase; the dotted lines show the
range of the minority spin component in the normal phase, which are
effectively the phase boundaries between the NM and the NP phase;
and the dash-dotted lines mark the range of the majority spin
component in the normal phase. The trap radius $\widetilde{r}$ is in
the units of the Thomas-Fermi radius for the corresponding
directions.}
\end{figure*}

\begin{figure*}[p]
\includegraphics{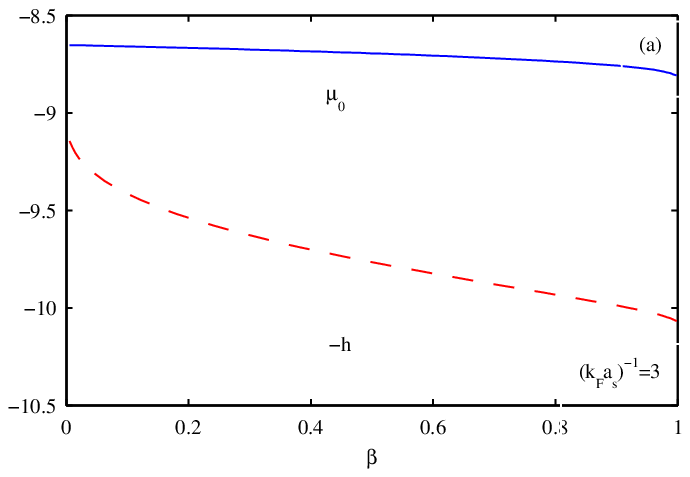} \includegraphics{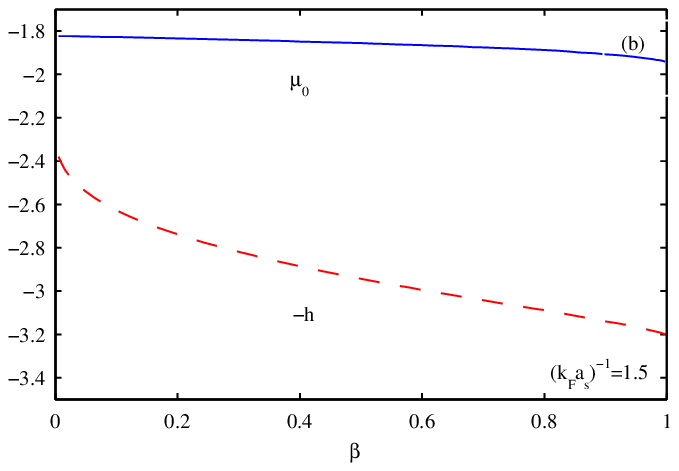} %
\includegraphics{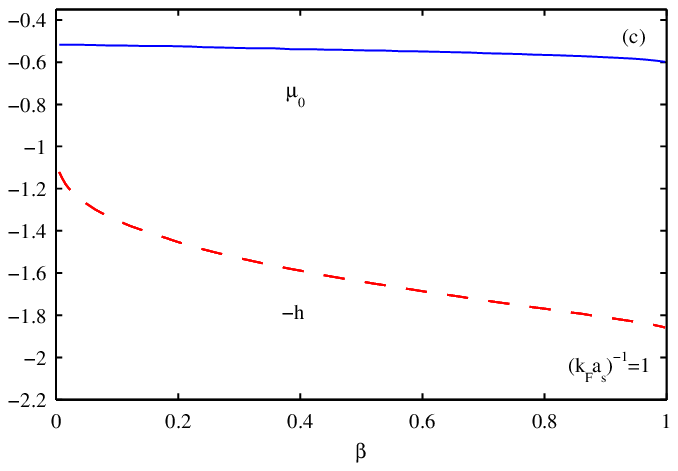} \includegraphics{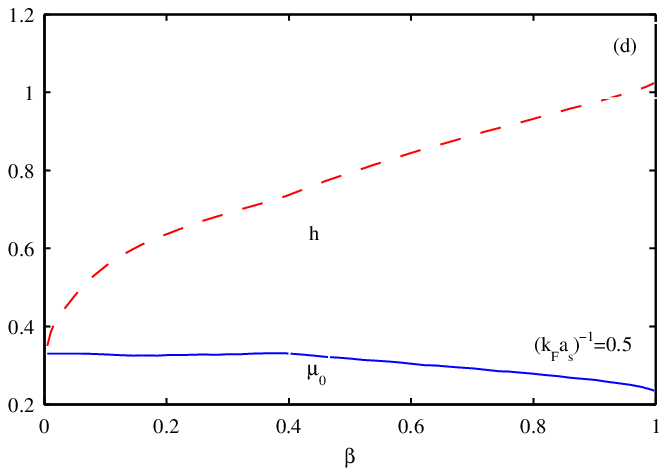} %
\includegraphics{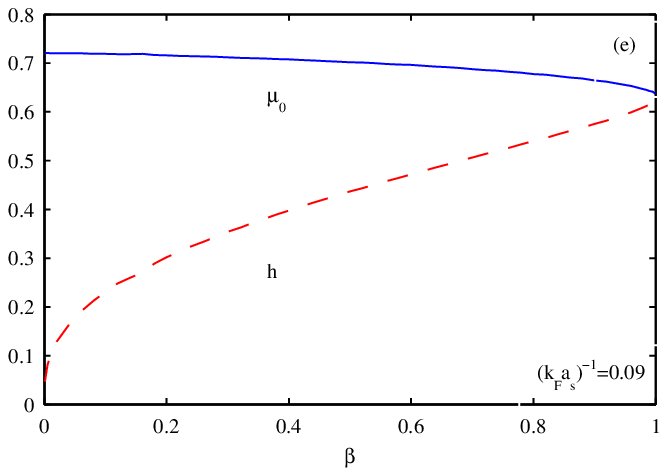} \includegraphics{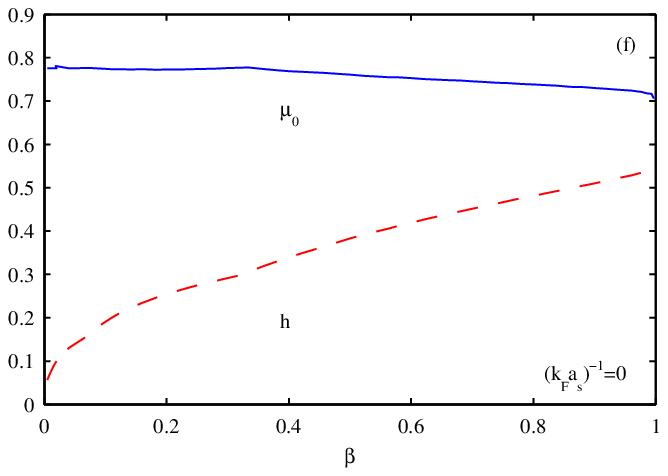} %
\includegraphics{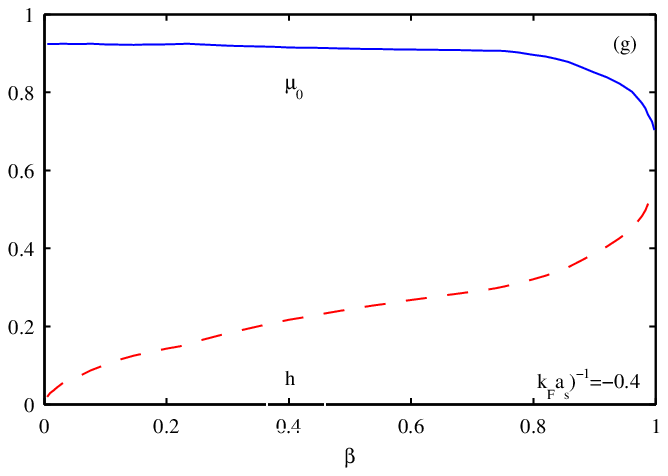} \includegraphics{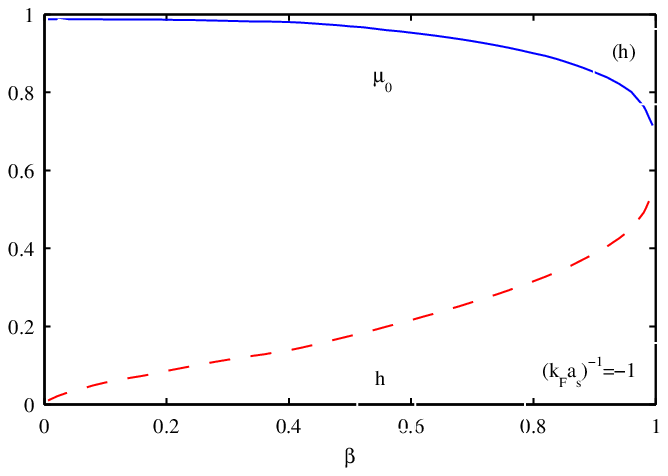}
\caption[Fig.2 ]{(Color Online) The chemical potential
$\protect\mu_0$ and the chemical potential difference $h$ as
functions of population imbalance at zero temperature for: (a)$(k_Fa_s)^{-1}=3$, (b) $%
(k_Fa_s)^{-1}=1.5$, (c) $(k_Fa_s)^{-1}=1$, (d) $(k_Fa_s)^{-1}=0.5$, (e) $%
(k_Fa_s)^{-1}=0.09$, (f) $(k_Fa_s)^{-1}=0$, (g) $(k_Fa_s)^{-1}=-0.4$, (h) $%
(k_Fa_s)^{-1}=-1$. The solid lines represent $\protect\mu_0/E_F$,
the average chemical potential at the center of the trap; and the
dashed lines
are $h/E_F$ ($-h/E_F$ in (a-c) for better comparison with $\protect\mu_0/E_F$%
), where the unit of energy $E_F$ is defined in Sec. II.}
\end{figure*}

\begin{figure*}[p]
\includegraphics{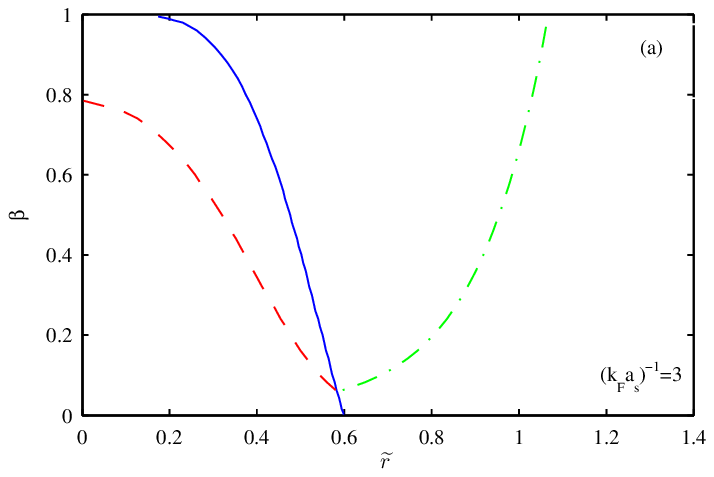} \includegraphics{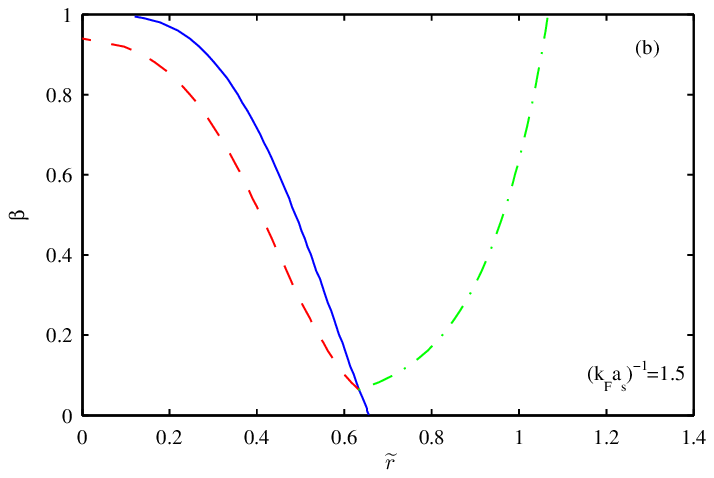} %
\includegraphics{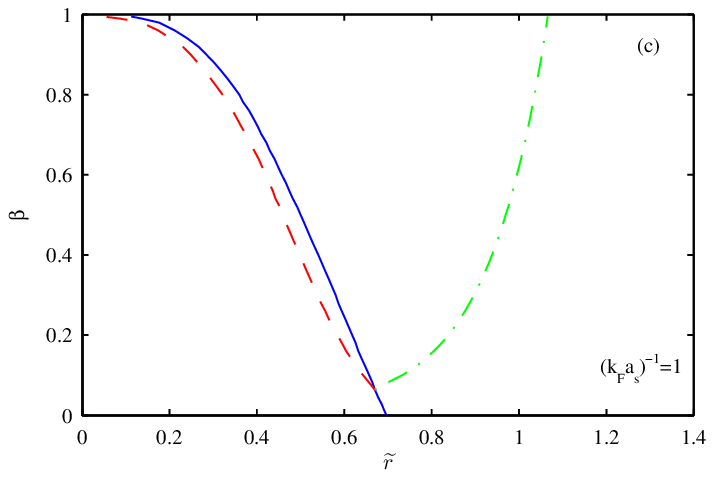} \includegraphics{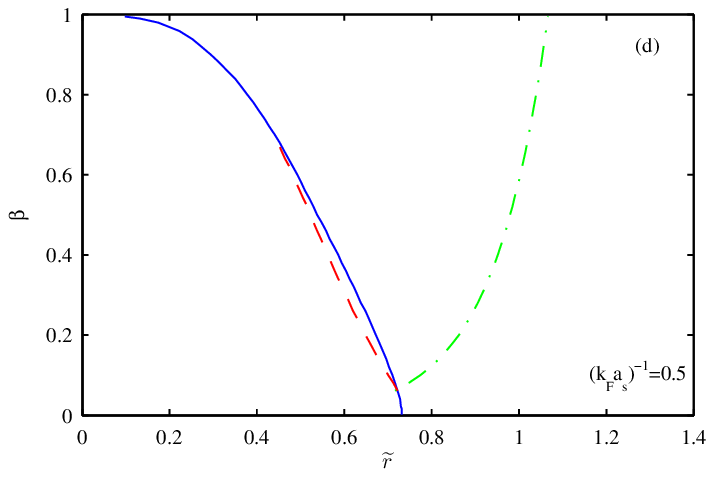} %
\includegraphics{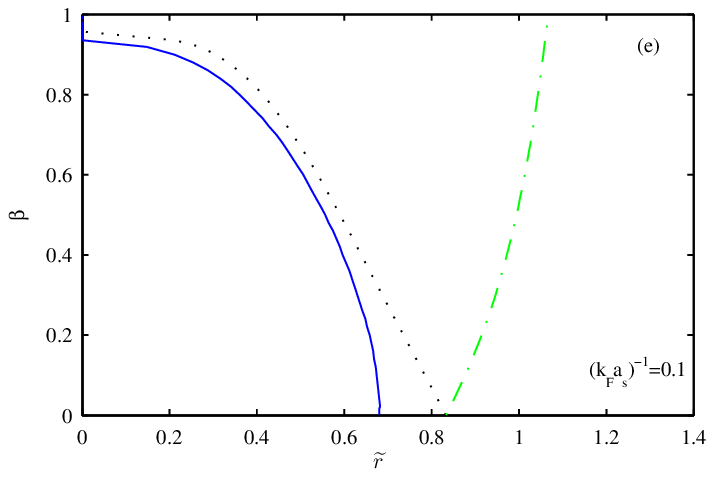} \includegraphics{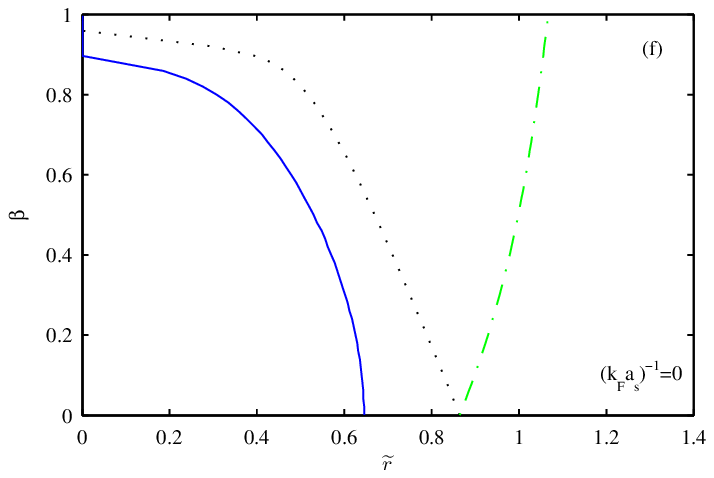} %
\includegraphics{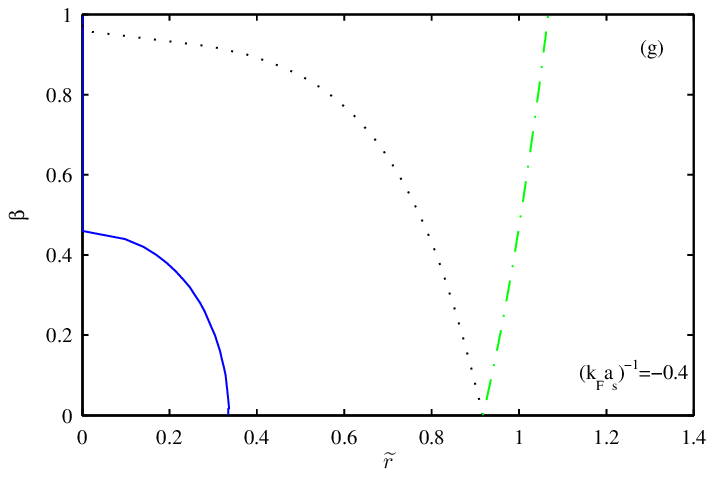} \includegraphics{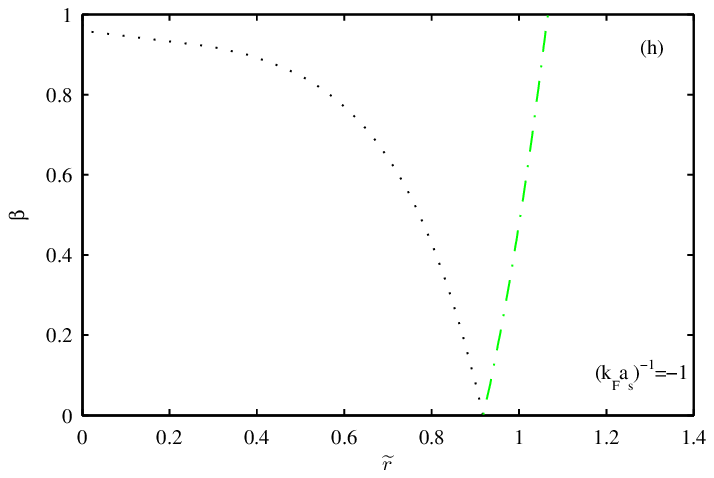}
\caption[Fig.3 ]{(Color Online) The finite temperature phase
boundaries of a
polarized fermion gas in a trap with (a)$(k_Fa_s)^{-1}=3$, (b) $%
(k_Fa_s)^{-1}=1.5$, (c) $(k_Fa_s)^{-1}=1$, (d) $(k_Fa_s)^{-1}=0.5$, (e) $%
(k_Fa_s)^{-1}=0.1$, (f) $(k_Fa_s)^{-1}=0$, (g) $(k_Fa_s)^{-1}=-0.4$, (h) $%
(k-Fa_s)^{-1}=-1$. The solid lines mark the phase boundaries between
the gapped and the gapless region; the dashed lines are the phase
boundaries between the gapped phases without a Fermi surface and the
ones that have; the dotted lines show the zero point of chemical
potential of the minority spin component in the normal phase; and
the dash-dotted lines show the zero point of the chemical potential
of the majority spin component in the normal
phase. The temperature is taken to be $T=0.22T_F$, which corresponds to a real temperature $%
T \sim 300nK$ for $2.7\times10^7$ fermionic $^6Li$ atoms in a cigar shaped trap with $%
\protect\omega_z\sim23Hz,\protect\omega_{\protect x} =
\omega_{\protect y} \sim110Hz$ \protect\cite{2}.}
\end{figure*}

\begin{figure*}[p]
\includegraphics{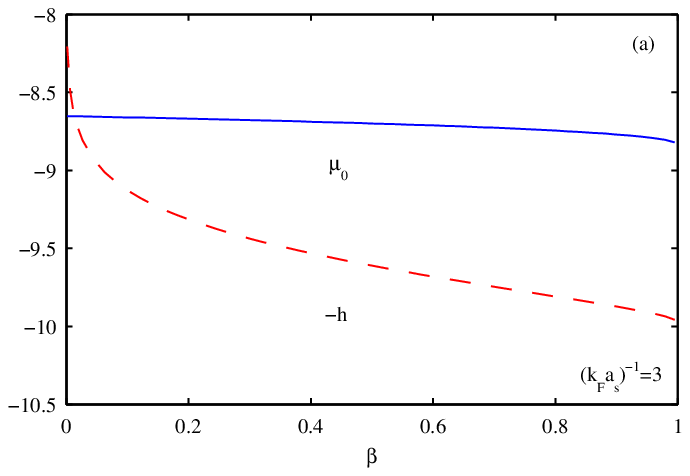} \includegraphics{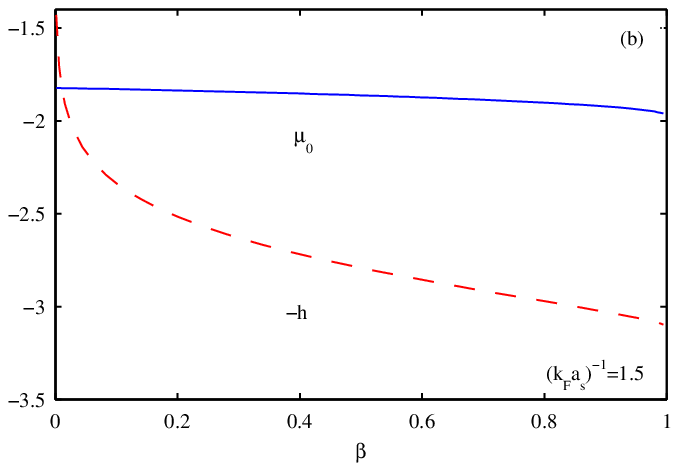} %
\includegraphics{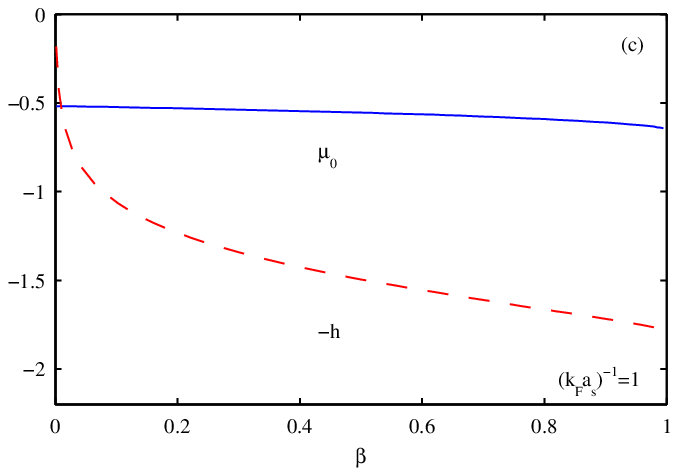} \includegraphics{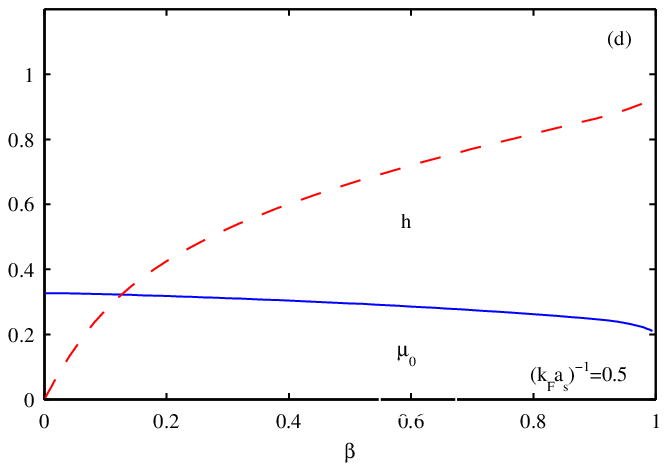} %
\includegraphics{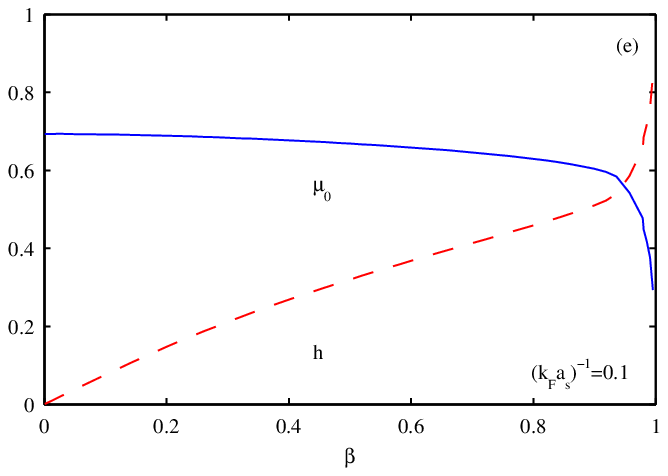} \includegraphics{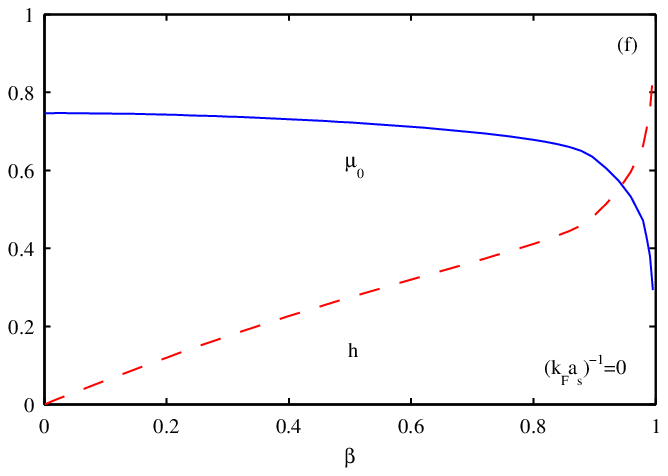} %
\includegraphics{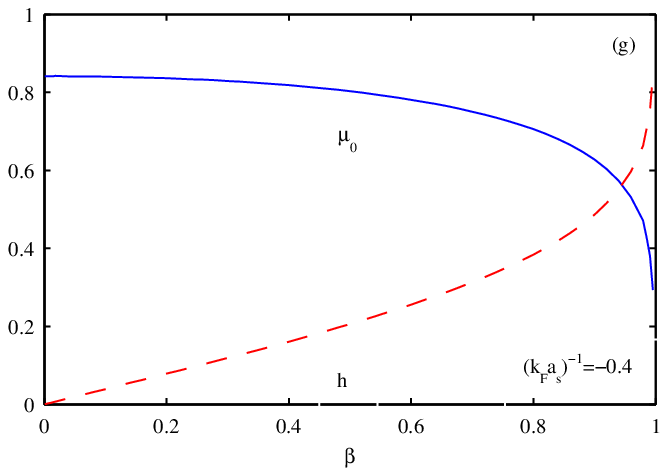} \includegraphics{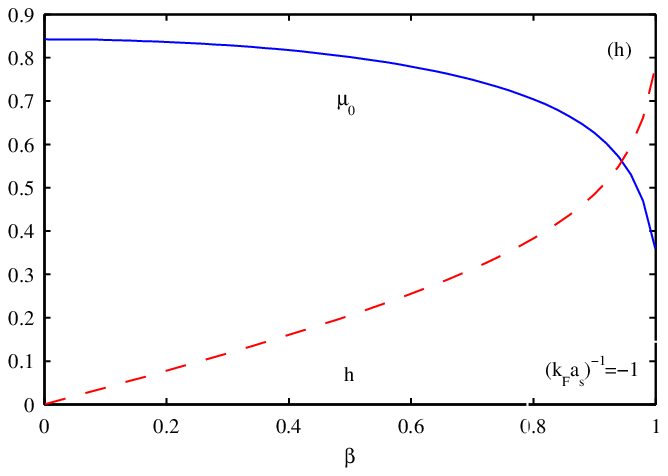}
\caption[Fig.4 ]{(Color Online) Same as Fig. 2, but $\protect\mu$
and $h$ are shown as functions of population imbalance at finite temperature ($T=0.22T_F$) with: (a)$%
(k_Fa_s)^{-1}=3$, (b) $(k_Fa_s)^{-1}=1.5$, (c) $(k_Fa_s)^{-1}=1$, (d) $%
(k_Fa_s)^{-1}=0.5$, (e) $(k_Fa_s)^{-1}=0.1$, (f) $(k_Fa_s)^{-1}=0$, (g) $%
(k_Fa_s)^{-1}=-0.4$, (h) $(k-Fa_s)^{-1}=-1$. The solid lines represent $%
\protect\mu/E_F$; and the dashed lines are $h/E_F$ }
\end{figure*}

Figure 1 shows the detailed phase distribution in the trap at
different interaction strengths $(k_{F}a_{s})^{-1}$ (corresponding
to different
magnetic field detunings). On the BEC side of the resonance (Fig.1(a), with $%
(k_{F}a_{s})^{-1}=3$) and at small but nonzero population
imbalances, the Fermi gas separates into three phases in the trap:
an SF core at the center, a BP1 phase in the middle, and an NP phase
at the edge of the
trap with only the majority spin component. As the imbalance parameter $%
\beta $ increases, the superfluid core becomes smaller until it vanishes at
a critical imbalance, beyond which only the BP1 phase and the normal phase
exist in the trap. This critical imbalance where the phase transition from
the SF phase to the BP1 phase occurs becomes greater towards the resonance,
while the parameter range of the BP1 phase shrinks (Fig. 1(b)(c)). At $%
(k_{F}a_{s})^{-1}=1$, the BP1 phase only exists at small population
imbalances in a slim region bordering the SF and the NP phase (Fig.
1(c)). Within our numerical resolution ($\delta\beta\sim\pm1\%$),
the BP1 phase disappears at roughly $(k_{F}a_{s})^{-1}\sim 0.5$, on
the BEC side of the resonance. Note that from numerical analysis, we
find
that the unpaired fermions in this BP1 phase are within the momentum range $%
0<\left| \mathbf{k}\right| <k_{+}$ (which implies $|\mu _{\mathbf{r}}|<\sqrt{%
h^{2}-|\Delta| ^{2}}$ in this region), where $k_{+}$ is given in the
previous section. In this BP1 phase, we therefore have the following
phase separation picture in the momentum space: the paired fermions
in the superfluid fill the outside momentum shell, while the
unpaired fermions of the majority spin
component occupy the states inside the Fermi ball with $\left| \mathbf{k}%
\right| <k_{+}$.

Once the BP1 phase disappears, the trap is left with only the SF
phase and the NP phase over a certain range of the interaction
strength (Fig. 1(d), with  $(k_{F}a_{s})^{-1}\sim 0.5$). This
continues till a new normal state,
the NM phase shows up at large population imbalances at roughly $%
(k_{F}a_{s})^{-1}\sim 0.1$ (Fig. 1(e)), where fermions of both spin
components show up in the normal gas, with different Fermi surfaces
for different spins. The range of the NM phase grows towards the
resonance, and at resonance or on the BCS side (Fig. 1(f)(g)(h)),
the Fermi gas typically phase separates into three regions at small
$\beta $: the SF phase, the NM phase and the NP phase. The
superfluid phase disappears at a critical population imbalance,
where the gas undergoes a phase transition from the superfluid phase
to the normal state. Qualitatively, this picture agrees pretty well
with the recent experimental findings \cite{2}, although
quantitatively, the mean-field type approximation at $T=0$ may
somewhat overestimate the critical population imbalance for the
disappearance of the SF core at the trap center.

To give more detailed information of this system, we also show in
Fig. 2 the chemical potential $\mu _{0}$ at the trap center and
the chemical potential difference $h$ as functions of the
population imbalance $\beta $ at the corresponding interaction
strengths $(k_{F}a_{s})^{-1}$. Note that the chemical potential
difference $h$ between the two spin components does not change
across the trap. The local chemical potential $\mu _{\mathbf{r}}$
changes, but given $\mu _{0}$ and the trap potential
$V(\mathbf{r)}$, it changes through the simple relation $\mu
_{\mathbf{r}}=\mu _{0}-V(\mathbf{r)} $ under the local density
approximation. With the information given in Fig. 2, we know the
local chemical potential $\mu _{\mathbf{r\sigma }}$ for each spin
component $\mathbf{\sigma }$, and it becomes straightforward to
calculate other properties of the system, such as the density
profiles.

\section{Phase Boundaries at Finite Temperature}

At finite temperature, we calculate the total gap ($\Delta $) in
the trap, from which we get the boundary between the gapped region
and the gapless region. As the order parameter for the superfluid
phase $\Delta _{s}$ is smaller than the total gap at finite
temperature, the boundary between the gapped and the gapless
regions serves as an upper bound for the superfluid phase
\cite{note4}. At finite temperature, due to the thermal
excitations, the atom density profiles change more smoothly as one
goes from the trap center to the edge, so one cannot easily use
the discontinuity of the density profiles to fix the boundary
between the BP1 phase and the SF phase, nor the boundary between
the different normal states (NM/NP). However, one can still look
at the changes of the Fermi surfaces in the momentum space, which
provides an unambiguous signal to determine the boundaries between
different phase/regions. We should caution, however, that these
boundaries do not necessarily correspond to sharp edges in the
atom density profiles. For instance, in the normal phase with
$\Delta =0$, when one of the chemical potentials, say $\mu
_{\mathbf{r\uparrow }}$, changes its sign at the trap edge, which
implies the disappearance of the Fermi surface for the spin up
atoms as one moves outside (under the local density
approximation), the density distribution of the spin up atoms does
not vanish at that point. Instead, it will follow an exponential
decay as the corresponding chemical potential turns negative. So,
although these boundaries do not represent sharp spatial edges of
the corresponding spin components, they indeed give a good
estimation of the ranges.

The results of our calculation at finite temperature are shown in
Fig. 3. We take the temperature $T=0.22T_{F}$, which corresponds to
$T\sim 300nK$ with the experimental parameters in Ref. \cite{2} (see
caption of Fig. 3). The phase diagrams are qualitatively similar to
those in Fig. 1, but there are several important differences at
finite T which deserve to be emphasized. First of all, on the BEC
side, the finite temperature BP1 region does not show up until a
critical population imbalance (see Fig. 3(a-d)). The basic reason
behind this feature is that at finite temperature, the population
imbalance can be carried by the quasiparticle excitations in the
conventional BCS\ superfluid state, as we have emphasized in
\cite{9c}. Therefore, different from the zero temperature case, the
conventional BCS\ state becomes partially polarized when there is a
population imbalance, which significantly relaxes the competition
between the population imbalance and the Cooper pairing, and which
makes the momentum-space phase separation between the paired state
and the excess fermions (the BP1 phase) unnecessary at small
population imbalances. Secondly, the range of the BP1 region changes
significantly from the zero temperature case. At $T=0.22T_{F}$, the
BP1 region disappears near $(k_{F}a_{s})^{-1}\sim 0.15$, where there
already exists an NM region in the phase diagram (see Fig. 3(d)(e)),
which initially appears near $(k_{F}a_{s})^{-1} \sim 0.3$ for small
population imbalance. Last but not least, at resonance or on the BCS
side, the critical population imbalance at which the gapped region
disappears at the center of the trap, which is an upper bound for
the actual critical imbalance bordering the SF phase and the
non-condensed pairs, becomes significantly smaller than the critical
population imbalance at zero temperature (see Fig. 3(f)(g)(h)). The
results above demonstrate that the phase boundaries, as well as the
evolution of the different phases as the field sweeps across the
crossover region, can be significantly affected by temperature. As
all the experiments are necessarily done at a finite temperature,
this suggests that it may be important to take the thermal effects
into account for a quantitative interpretation of the experimental
data.

At finite temperature, we also calculate the chemical potentials $\mu _{%
\mathbf{r\sigma }}$ as functions of the population imbalance
$\beta $ at various interaction strengths $(k_{F}a_{s})^{-1}$, in
order to provide detailed information of this system. The results
are shown in Fig. 4. The properties can be directly read from the
figure, and as the main features there are qualitatively similar
to those in Fig. 2 for the zero temperature case, we neglect the
detailed discussion here.

\section{Summary}

In summary, we provide detailed calculations of the phase diagrams
for a trapped Fermi gas with population imbalance over the entire
BCS-BEC crossover region at zero and at finite temperature. The
calculation is done with the self-consistent $G_{0}G$ diagram
scheme. We also emphasize the importance of using the universal
dimensionless equations. The properties of the system then only
depend on three dimensionless parameters, from which we can
calculate the universal phase diagrams valid for different atom
species with different atom numbers under various trap
configurations. The main results of our calculation are shown in
Fig. 1 and 3, with their prominent features discussed in detail in
the corresponding sections.

We thank Martin Zwierlein for helpful discussions. This work was
supported by the NSF awards (0431476), the ARDA under ARO
contracts, and the A. P. Sloan Fellowship.


\begin{thebibliography}{99}
\bibitem{note1}  Although there is an indication that the FFLO state might
exist near resonance in a narrow boundary region in the
trap\cite{9b}, we have not yet found evidence for the existence of
such a state which is stable in the strongly interacting region
through numerical search within the local density approximation.
\bibitem{note2}  If the trap has anharmonicity which is a constant expressed
in the unit of the Thomas-Fermi radius (see M. W. Zwierlein, W. Ketterle,
cond-mat/0603489), we will still have the same universality as discussed in
the text.
\bibitem{note3}  Note that this state is also called the polarized
(magnetized) superfluid state in the literature \cite{8a}.
However, at finite temperature, the normal BCS\ superfluid state
also has finite polarization as emphasized in Ref. \cite{9c}. To
distinguish between these different kinds of superfluid phases, we
call the non-BCS superfluid as the BP1 phase (abbreviation for the
breached pair phase with one Fermi surface), similar to the
terminology of Ref. \cite{7}. The breached pair state was studied
in Ref. \cite{7} on the BCS\ side, where it has two coexisting
Fermi surfaces (BP2 phase). Other than that difference, the
properties of the BP1 and BP2 phases are pretty much the same. For
instance, they are both superfluid states with gapless fermionic
excitations, and are both homogeneous in the real space while
phase separated in the momentum space.
\bibitem{note4}  To exactly fix the phase boundary for the superfluid phase
at finite $T$, one needs to derive the dispersion relation for the
pair excitations to split $\Delta $ into $\Delta _{s}$ and $\Delta
_{pg}$. This dispersion relation is known for the equal-population
case \cite{10}, but has not yet been derived for the case with
population imbalance within the self-consistent $G_{0}G$ scheme.
However, for the calculation of the total gap $\Delta $ and the atom
density profiles, one does not need to fix that dispersion relation
\cite{9c}.
\end{thebibliography}
\end{document}